\documentstyle[multicol,aps]{revtex}
\begin{document}
\title{Stochastic Aggregation: Rate Equations Approach}
\author{P.~L.~Krapivsky$\dag$ and E.~Ben-Naim$\ddag$}
\address{$\dag$Center for Polymer Studies and Department of Physics,
Boston University, Boston, MA 02215, USA}
\address{$\ddag$Theoretical Division and Center for Nonlinear Studies,
Los Alamos National Laboratory, Los Alamos, NM 87545, USA}
\maketitle
\begin{abstract}
  We investigate a class of {\em stochastic} aggregation processes
  involving two types of clusters: active and passive. The mass
  distribution is obtained analytically for several aggregation rates.
  When the aggregation rate is constant, we find that the mass
  distribution of passive clusters decays algebraically. Furthermore,
  the entire range of acceptable decay exponents is possible. For
  aggregation rates proportional to the cluster masses, we find that
  gelation is suppressed. In this case, the tail of the mass
  distribution decays exponentially for large masses, and as a power
  law over an intermediate size range.

\end{abstract}
\begin{multicols}{2}

\section{Introduction}

Aggregation, coagulation, or agglomeration are kinetic phenomena in
which clusters bond irreversibly to form clusters of indefinitely
growing mass. A detailed model of merging of two clusters into a single
cluster should incorporate the mass, position, velocity (or diffusion
rate), and even geometrical characteristics of each cluster, together
with the precise merging mechanism. Such a detailed description is well
beyond theoretical analysis. The most natural approach {\em assumes}
that the merging of two clusters of masses $x$ and $y$ into a cluster of
mass $x+y$ occurs with a rate kernel $K(x,y)$.  The cluster densities
then evolve according to the Smoluchowski coagulation
equations\cite{smol}.  This theoretical framework
\cite{smol,chan,drake,lt,ev} has found numerous applications in physical
chemistry\cite{sein}, as well as in astronomy\cite{pitman} and
mathematics\cite{aldous}.

Characterizing a cluster by a single parameter, its mass, may indeed
be a drastic oversimplification.  For instance, two clusters with a
similar mass can be substantially different.  Additionally, the
character of a cluster may be altered every time it undergoes
aggregation, and for example, it may become less likely to participate
in further aggregation events. In this paper, we investigate
situations where active clusters may become passive after merging
events. A concrete example where active and passive clusters coexist
are multi-phase coarsening processes in one dimension. Indeed, upon
merging, domain walls may remain active or become passive depending on
the phase of their neighboring domains\cite{domain,domain1,Ledoussal}.
Another example is polymerizations of linear polymers. Here, if the
end monomers can be chemically active or inert, newly produced
polymers may become passive. Scaling properties and numerical studies
of several stochastic aggregation problems will be presented in
\cite{next}.

Specifically, we consider a class of models where newly-formed
clusters remain active with a fixed probability $p$; otherwise, they
become passive in the sense that they never aggregate again. As the
fate of newly-formed clusters is determined stochastically, we term
this process {\em stochastic aggregation}.  Given a fixed probability
for active clusters to become passive, all clusters eventually become
passive, and the system freezes. Our goal is to determine the time
dependent distributions of active and passive clusters, as well as the
final mass distribution of passive clusters.

Within the standard framework\cite{smol,chan}, it has long been
recognized that the Smoluchowski master equations are tractable for
three particular kernels: $K(x,y)=1$, $x+y$, and $xy$\cite{drake}. Exact
results have also been found for linear combinations of the above
solvable kernels \cite{spouge,treat}, and for one nonlinear
interpolation between the constant and sum kernels \cite{leyvraz}. As
will be shown below, stochastic aggregation is solvable for the three
classical kernels.

For instance, in the simplest case of mass independent aggregation
rates, we find that $P_k$, the final mass distribution of passive
clusters, is scale free, i.e, $P_k\sim k^{-\gamma}$. Since
$\gamma=2/p$, the entire range of decay exponents $2<\gamma<\infty$ is
realized by tuning $0<p<1$. Similar behavior emerges in a dual
fragmentation process where newly-formed fragments may become passive
\cite{kgb}. The sum and the product kernels behave quite similarly as
the final mass distribution is suppressed exponentially for large
masses. Nevertheless, over an intermediate size range, an algebraic
decay with a fixed decay exponent may occur. The time dependent
behavior is governed by an algebraically growing size scale, and it
can be cast as either a scaling or a scaling-like form.  In general,
the growth law for this scale is not universal as it depends on the
parameter $p$.

The rest of this paper is organized as follows.  In Sec.~II, we define
the model and outline the general framework. Exact solutions of the
master equations for the cases of constant, product, and sum kernels are
presented in Secs.~III, IV, and V, respectively. The results are
summarized and discussed in Sec.~VI.

\section{The model}

Consider the initial conditions where all clusters are active and
have the same mass, which can be set to unity without loss of
generality. Then, active monomers aggregate to form dimers, and so
on. After an aggregation event, the newborn cluster remains active
with probability $p$, or becomes passive with probability $q=1-p$. In
the latter case, the cluster does not undergo further aggregation.

Let $K(i,j)$ be the rate by which clusters of mass $i$ and clusters of
mass $j$ aggregate, and let $A_k$ ($P_k$) denote active (passive)
clusters of mass $k$.  The stochastic aggregation process can be
symbolically written as
\begin{eqnarray}
\label{symb}
A_i+A_j\buildrel pK(i,j) \over \longrightarrow& A_{i+j},\quad
A_i+A_j\buildrel qK(i,j) \over \longrightarrow&P_{i+j}.
\end{eqnarray}
We use the notations $A_k(t)$ and $P_k(t)$ to denote the density
distributions of active and passive clusters at time $t$.  The master
equations describing the time evolution of the system read
\begin{eqnarray}
\label{me}
{d A_k\over dt}&=&{p\over 2}\sum_{i+j=k}K(i,j)A_i A_j
-A_k\sum_{j\geq 1} K(k,j) A_j,\nonumber\\
{d P_k\over dt}&=&{q\over 2}\sum_{i+j=k}K(i,j)A_iA_j.
\end{eqnarray}
The convolution terms are proportional to $p$ and $q$, respectively,
reflecting the probabilities of remaining active and becoming passive
after each aggregation event. The master equations are subject to 
monodisperse initial conditions
\begin{equation}
\label{ic}
A_k(0)=\delta_{k,1},\qquad P_k(0)=0.
\end{equation}
One may verify that the overall mass is conserved,  
i.e, $\sum_{k\geq 1} k[A_k(t)+P_k(t)]=1$.

We are interested in the temporal behavior of the mass distributions
of both active and passive clusters, and in the final distribution of
passive clusters.  An important feature of the model is that the
number densities of active and passive clusters are ultimately
related throughout the evolution.  Indeed, in each aggregation event,
the total number of active clusters is reduced by $1$ with probability
$p$, or by $2$ with probability $q$; hence on average, it reduces by
$1+q$.  Simultaneously, each aggregation event increases the total
number of passive clusters by $q$ on average.  Thus, the following
properly weighted combination of the number densities
$A(t)=\sum_{k\geq 1} A_k(t)$ and $P(t)=\sum_{k\geq 1} P_k(t)$ is
conserved
\begin{equation}
\label{pqa}
qA(t)+(1+q)P(t)={\rm const}.
\end{equation}
This second conservation law follows from the rate equations as well.
However, it holds generally for stochastic aggregation processes, even
when the master equations (\ref{me}) do not apply.

Assuming that initially there were no passive clusters, $P(0)=0$, and
setting the initial cluster density to one, $A(0)=1$, we see that
Eq.~(\ref{pqa}) gives a neat expression for the final number density
of passive clusters
\begin{equation}
\label{pinf}
P(\infty)={q\over 1+q}.
\end{equation}
Additionally, mass conservation implies $\sum_k kP_k(\infty)=1$, and
therefore the average mass of a passive cluster is $(1+q)/q$.  A
$q^{-1}$ divergence occurs in the $q\to 0$ limit, corresponding to
traditional aggregation.  Remarkably, the final density (or
equivalently, the average mass) is independent of the details of the
model including the aggregation kernel, the clusters' transport
mechanism, or the system's dimensionality.

In what follows, we solve the master equations for three different
kernels: constant $K(i,j)=2$, product $K(i,j)=ij$, and sum
$K(i,j)=i+j$.  Although we restrict our attention to monodisperse
initial conditions, our techniques apply to arbitrary initial
conditions. Furthermore, the nature of the solutions extends to
compact initial distributions.

\section{Constant Kernel}

The constant kernel is the most widely used one.  It has been applied to
diffusion limited coalescence, for example. In traditional aggregation
with $K(i,j)={\rm const}$, the typical mass grows linearly with time,
and the mass distribution is exponential. We shall see below that the
latter assertion also applies for stochastic aggregation.

We conveniently set $K(i,j)=2$, and the master equations read
\begin{eqnarray}
\label{meconst}
{d A_k\over dt}&=&p\sum_{i+j=k}A_iA_j-2AA_k,\nonumber\\
{d P_k\over dt}&=&q\sum_{i+j=k}A_iA_j.
\end{eqnarray}
The number densities of active and passive clusters can be obtained
directly. The reaction proceeds through two channels: $A+A\to A$ with
rate $p$ and $A+A\to P$ with rate $q$, and therefore, the number
densities obey 
\begin{equation}
\label{apt}
{d A\over dt}=-(1+q)A^2, \qquad {d P\over dt}=qA^2.
\end{equation}
Solving these equations subject to the initial conditions $A(0)=1$ and
$P(0)=0$ gives
\begin{eqnarray}
\label{rp}
A(t)= {1\over 1+(1+q)t},\qquad P(t)={qt\over 1+(1+q)t}.
\end{eqnarray}
Indeed, the relation (\ref{pqa}) between $A$ and $P$ holds, and the
final density $P(\infty)=q/(1+q)$ is recovered. Additionally, the
density of active clusters decays algebraically with a universal
exponent: $A\sim t^{-1}$.

One can also study the mass density of active clusters,
$M(t)=\sum_{k\geq 1} kA_k(t)$.  This quantity has a nice probabilistic
interpretation: it equals the survival probability of an active
monomer, initially present in the system. In other words, it is the
probability that such monomer belongs to an active cluster at time
$t$.  {}From Eq.~(\ref{meconst}) we find that the mass density evolves
according to ${d\over dt}M=-2qAM$. Using $A(t)$ from Eq.~(\ref{rp})
and the initial condition $M(0)=1$, we integrate the rate equation to
find
\begin{equation}
\label{mut}
M(t)=[1+(1+q)t]^{-{2q\over 1+q}}.
\end{equation}
Hence, the decay of the total mass density and the growth of the
average cluster size exhibit non universal behavior as they depend on
the parameter $q$: $M\sim t^{-2q/(1+q)}$ and $\langle k\rangle=M/A\sim
t^{p/(1+q)}$, respectively.  Note also that the linear average mass
growth, $\langle k\rangle \sim t$, is recovered for the traditional
aggregation process $p=1$. Similarly, the monomer density
$A_1(t)=[1+(1+q)t]^{-2/(1+q)}$, obtained from ${d\over dt}A_1=-2AA_1$,
exhibits an algebraic decay with a $q$-dependent exponent. This
quantity is the probability that a monomer avoids aggregation events up to time
$t$.

We solve for the mass distribution of active clusters first. To
this end, we transform the distribution $A_k$ and introduce a
modified time variable $\tau$ as follows
\begin{eqnarray}
\label{atau}
{\cal A}_k&=&A_k\,\exp\left[2\int_0^t dt'\,A(t')\right],\nonumber \\
\tau&=&p\int_0^t dt_1\,\exp\left[-2\int_0^{t_1} dt_2\,A(t_2)\right].
\end{eqnarray}
The modified time variable can be written explicitly,
$\tau=1-[1+(1+q)t]^{-{p\over 1+q}}$. In terms of this time variable
one has $A=(1-\tau)^{1+q\over p}$, and ${\cal
  A}_k=A_k(1-\tau)^{-2/p}$.  The above transformations reduce
Eq.~(\ref{meconst}) to
\begin{equation}
\label{aktild}
{d {\cal A}_k\over d\tau}=\sum_{i+j=k} {\cal A}_i {\cal A}_j.
\end{equation}
These equations can be solved by the generating functions
technique. Indeed, ${\cal A}(z,\tau)=\sum_{k\geq 1} e^{k z} {\cal
A}_k(\tau)$ obeys
\begin{equation}
\label{azt}
{\partial {\cal A}(z,\tau)\over \partial \tau}={\cal A}^2(z,\tau).
\end{equation}
The solution is given by ${\cal A}(z,\tau)=A_0(z) [1-\tau
A_0(z)]^{-1}$, with $A_0(z)\equiv A(z,0)={\cal A}(z,0)$
the initial generating function. For the  monodisperse
initial conditions (\ref{ic}), ${\cal A}_0(z)=e^z$, and we find
${\cal A}_k=\tau^{k-1}$. Thus, the original mass distribution reads
\begin{equation}
\label{ak}
A_k(\tau)=(1-\tau)^{2/p}\,\tau^{k-1},
\end{equation}
i.e., it remains exponential throughout the evolution.  
Although we present this solution as a function of $\tau$, it
is an explicit solution as $\tau(t)$ is known.

The passive cluster distribution can be obtained by integrating the
corresponding master equation. Again, it is more convenient to use the
modified time variable as Eq.~(\ref{meconst}) simplifies to ${d\over
  d\tau}P_k={q\over p}\,(k-1)(1-\tau)^{2/p}\tau^{k-2}$. Integrating
over the modified time gives
\begin{equation}
\label{pa}
P_k(\tau)={q\over p}\,(k-1)\int_0^\tau dx\,(1-x)^{2/p}\,x^{k-2}.
\end{equation}
The final distribution is obtained by setting $\tau=1$. It can be
expresses in terms of the Euler Gamma function,
\begin{equation}
\label{pfinal}
P_k(\infty)={q\over p}\,{\Gamma(1+2/p)\,\Gamma(k)\over \Gamma(k+2/p)}.
\end{equation}
The large mass tail of the distribution is therefore suppressed
algebraically according to
\begin{equation}
\label{pkinf}
P_k(\infty)\simeq qp^{-1}\Gamma(1+2/p)\,k^{-2/p}.
\end{equation}
Interestingly, given the algebraic decay $P_k\sim k^{-\gamma}$ as
$k\to\infty$, mass conservation restricts the exponent range to
$\gamma>2$. In our case $\gamma=2/p$, and since $0<p<1$, the entire
range of acceptable exponents emerges by tuning the only control
parameter $p$. This behavior is reminiscent of a related stochastic
fragmentation process where newly-formed fragments may turn
stable\cite{kgb}.

Despite their different limiting behaviors, both mass distributions
obey the same scaling form. This reflects the nature of the process:
the distributions are coupled, and the ``activity'' in the system 
involves masses of the order of the characteristic mass. To obtain the
scaling forms, we consider the limit $k\to\infty$ and $t\to\infty$,
with the scaling variable $\xi=k(1-\tau)$ kept finite.  In this limit,
Eqs.~(\ref{ak}) and (\ref{pa}) acquire the following scaling forms
\begin{eqnarray}
\label{scalconst}
A_k(t)\sim t^{-{2\over 1+q}} F(\xi),\qquad
P_k(t)\sim t^{-{2\over 1+q}} G(\xi)
\end{eqnarray}
with the scaling functions
\begin{eqnarray}
\label{psiphi}
F(\xi)=\exp(-\xi), \qquad
G(\xi)=\xi^{-2/p}\Gamma(1+2/p,\xi).
\end{eqnarray}
Here, $\Gamma(a,\xi)=\int_\xi^{\infty}dx\, x^{a-1}\,e^{-x}$ is the
incomplete Gamma function. The scaling variable $\xi=k/k^*$ contains
the characteristic mass $k^*=(1-\tau)^{-1}=\langle k\rangle\sim t^{p/(1+q)}$.
At time $t$, the mass of passive clusters being produced is on the
order $k^*$. Masses larger than this scale are exponentially rare as
$G(\xi)\propto F(\xi)=e^{-\xi}$. On the other hand, masses
smaller than this scale have already turned passive. This is
manifested by the small argument divergence $G(\xi)\sim
\xi^{-2/p}$, which leads to the {\rm time independent} algebraic
distribution (\ref{pkinf}).

In fact, the scaling functions are unique.  When the initial mass is
finite, the scaling exponents and consequently, the scaling form
(\ref{scalconst}) are dictated by the rate equations for the number
and mass densities.  Substituting this scaling form into the rate
equations (\ref{meconst}) yields an integro-differential equation for
the scaling function $\int_0^\xi d\eta\,F(\eta)F(\xi-\eta)+\xi
F'(\xi)=0$.  This equation can be solved via the Laplace transform,
and as long as the first two moments of $F$ are finite, the solution
is a simple exponential. A similar analysis can also be carried for
the second scaling function.  We conclude that the above scaling
behavior holds in general for compact initial distributions.

\section{Product Kernel}

The aggregation process with a reaction rate proportional to the mass
of both reactants, $K(i,j)=ij$, has been applied to polymerizations of
branched polymers \cite{drake}, and to random graphs\cite{bal}.  In
this case, the system exhibits a phase transition.
Specifically, the mass condenses into an infinite cluster. In the
following, we show that  gelation does not occur when there is a
finite probability for clusters to turn passive.

In this case, the master equations (\ref{me}) read
\begin{eqnarray}
\label{meprod}
{d A_k\over dt}&=&{p\over 2}\sum_{i+j=k}ijA_iA_j-kA_kM,\nonumber\\
{d P_k\over dt}&=&{q\over 2}\sum_{i+j=k}ijA_iA_j.
\end{eqnarray}
Here as well $M(t)=\sum_{k\ge 1} kA_k(t)$ is the mass density of
active clusters.  Let us start with analysis of the active cluster
distribution.  To facilitate the solution, we again employ the
generating functions technique.  Specifically, the generating function
of the sequence $kA_k$, ${\cal A}(z,t)=\sum_{k\geq 1} kA_k(t)e^{kz}$,
evolves according to
\begin{equation}
\label{Rzt}
{\partial {\cal A}\over \partial t}=(p{\cal A}-M)
{\partial {\cal A}\over \partial z}.
\end{equation}
It is useful to re-write partial derivatives as Jacobians: ${\partial
  {\cal A}\over \partial t}= {\partial ({\cal A}, z)\over \partial
  (t, z)}$ and ${\partial {\cal A}\over \partial z}= {\partial
  ({\cal A}, t)\over \partial (z, t)}$.  Substituting these
Jacobians into Eq.~(\ref{Rzt}), and using the identity ${\partial
  z\over \partial t}= {\partial (z, {\cal A})\over \partial (t,
  {\cal A})}$, we arrive at
\begin{equation}
\label{zRt}
{\partial z\over \partial t}=M-p{\cal A}.
\end{equation}
This equation suggests that one should solve for $z({\cal A},t)$.
Integrating Eq.~(\ref{zRt}) gives a solution up to an arbitrary constant
$F({\cal A})$: $z({\cal A},t)=\int_0^t dt'\,M(t')-pt{\cal
  A}+F({\cal A})$.  The initial conditions (\ref{ic}) read ${\cal
  A}_0(z)=e^z$ or alternatively, $F({\cal A})=\ln {\cal A}$, thereby
leading to
\begin{equation}
\label{zA}
z({\cal A},t)=\int_0^t dt'\,M(t')-pt{\cal A}+\ln {\cal A}.
\end{equation}

Therefore, the problem is now reduced to evaluation of the mass
density of active clusters $M(t)$.  Substituting ${\cal 
  A}(z=0,t)=M(t)$ into Eq.~(\ref{zA}) gives 
\begin{equation}
\label{mu}
\int_0^t dt'\,M(t')-ptM(t)+\ln M(t)=0.
\end{equation}
Differentiating this equation leads to $(ptM-1){dM\over
  dt}=qM^2$.  Rather than solve for $M(t)$, one can obtain an
explicit solution for $t(M)$. Indeed, the first order differential
equation $qM^2\,{dt\over dM}-ptM=-1$ can be easily solved by a
number of techniques, e.g., by variation of parameters, to give
\begin{equation}
\label{t}
t=M^{-1}-M^{p/q}.
\end{equation}
As a check of self-consistency, we confirm that this expression 
agrees with the first order Taylor series for the mass
$M(t)\cong 1-qt$, implied by the master equation (\ref{meprod}).
Equation (\ref{t}) in principle gives the mass density $M(t)$.  To
determine the number density $A(t)$, we sum up Eqs.~(\ref{meprod}) and
find ${d\over dt} A=-{1+q\over 2} \,M^2$.  Since we have an explicit
expression for $t(M)$ rather than $M(t)$, we 
treat $A$ as a function of $M$ as well. The density obeys ${d A\over
  dM}={1+q\over 2}\left[1+{p\over q}\,M^{1/q}\right]$, which can
be integrated to yield
\begin{equation}
\label{Amu}
A(M)={M\over 2}\left[1+q+pM^{1/q}\right].
\end{equation}
Equations (\ref{t}) and (\ref{Amu}) show that both the number and mass
densities of active clusters decay similarly in the long time limit,
\begin{equation}
\label{Amut}
A(t)\simeq {1+q\over 2}\,t^{-1},\qquad M(t)\simeq t^{-1}.
\end{equation}
The decay exponent is universal and identical to that of the constant
kernel model. In that case, however, only the number density decayed
as $t^{-1}$. Eq.~(\ref{Amut}) also offers insight to the nature of the
process as the average mass approaches a constant $\langle
k\rangle=M/A\to 2/(1+q)$. One can verify that this quantity
approaches unity in the limit $q\to 1$ when the active mass distribution
contains only monomers.

To complete the solution, we need to determine $A_k$. Note that
Eq.~(\ref{zA}) can be re-written as ${\cal A}\,e^{-pt{\cal
    A}}=\exp\left[z-\int_0^t dt'\,M(t')\right]$.  The Lagrange
inversion formula is quite handy here \cite{wilf}. It states that the
solution for the equation $ve^{-v}=u$ can be expressed in terms of the
series $v=\sum_{n\ge 1} {n^{n-1}\over n!}u^n$. Applying it to the
above equation yields
\begin{equation}
\label{Akt}
A_k(t)={(kpt)^{k-1}\over k\cdot k!}
\exp\left[-k\int_0^t dt'\,M(t')\right].
\end{equation}
We again express $A_k$ as a function of $M$ rather than $t$.
The integral  can be read from Eq.~(\ref{mu}), while
the time $t(M)$ can be read from Eq.~(\ref{t}). Substituting the
corresponding expressions  yields the active cluster mass distribution
in terms of the variable $\nu=M^{1/q}$
\begin{equation}
\label{Akmu}
A_k(\nu)=\nu^q\,{[kp(1-\nu)]^{k-1}\over k\cdot k!}\,e^{-kp(1-\nu)}.
\end{equation}
In the large mass limit, the leading behavior of this mass distribution
is given by
\begin{equation}
\label{Lambda}
A_k(\nu)\simeq {\nu^q\over \sqrt{2\pi p^2(1-\nu)^2}}
\,k^{-5/2}\,e^{-\Lambda(\nu,q) k},
\end{equation}
with the decay constant
$\Lambda(\nu,q)=q\nu-[q+\ln(1-q)]-[\nu+\ln(1-\nu)]$. This constant must
be positive and this is confirmed by the expansion
$\Lambda(\nu,q)=q\nu+\sum_{n\geq 2} n^{-1}(q^n+\nu^n)$.  Therefore, the
mass distribution decays exponentially with mass. This in particular
implies that no gelation occurs. However, over an intermediate range
$k\ll \Lambda^{-1}(\nu,q)$, the mass distribution decays algebraically,
$A_k\sim t^{-1} k^{-5/2}$. Furthermore, the monomer density equals
$A_1=M\,e^{-p+pM^{1/q}}$, and in the long time limit $A_1\simeq e^{-p}
t^{-1}$.

The asymptotic behavior is of special interest. In the limit
$k\to\infty$ and $t\to\infty$ with the scaling variable $\xi=kq\nu$
kept finite, the mass distribution of active clusters can be re-casted
into a ``scaling-like'' form
\begin{eqnarray}
\label{qscl}
A_k(t)\sim e^{-\Lambda(q) k}\,k^{-5/2}\,t^{-1}\,F(\xi),
\end{eqnarray}
with $\Lambda(q)=-q-\ln(1-q)$.  The scaling function is simply
exponential, $F(\xi)=e^{-\xi}$.  The above differs from an ordinary
scaling form in that it is cannot be reduced from two variables $(k,t)$
to one rescaled variable. Furthermore, the exponential factor
$e^{-\Lambda k}$ indicates that the mass distribution is appreciable
over a fixed mass range of the order $1/\Lambda(q)$. This range is
significant only in the (traditional aggregation) limit $q\to 0$. In
this case, a universal algebraic decay $A_k\sim t^{-1}k^{-5/2}$ is
realized in the intermediate mass range $k\ll q^{-2}$. Despite this, it
is interesting that the distribution is characterized by a scaling-like
factor, with a scaling variable $\xi=k/k^*$.  Note that the
characteristic mass grows with time, $k^*\sim q^{-1}t^{1/q}$, although
the average mass $\langle k\rangle$ remains finite.

Turning to passive clusters, we first note that their number density
can be obtained using $A(M)$ of Eq.~(\ref{Amu}) and the
conservation law (\ref{pqa}) 
\begin{equation}
\label{Pmu}
P(M)={q\over 1+q}-{qM\over 2}\left[1+{p\over 1+q}M^{1/q}\right].
\end{equation}
Alternatively, this density can be obtained by integrating ${d\over
  dt}P={1\over 2}qM^2$.  While the final number density (\ref{pinf})
is universal, the mass distribution of passive clusters is model
dependent. In the previous case of the constant kernel, we evaluated
the convolution term in the master equation explicitly. In the present
case, it is more convenient to circumvent this sum by noting that it
appears in both evolution equations. The evolution equation for
$P_k(t)$ can therefore be simplified to ${d\over dt}P_k(t)={q\over
  p}[{d\over dt}A_k+kA_kM]$ for $k\ge 2$.  Again, we  present
the mass distribution in terms of $\nu=M^{1/q}$
\begin{equation}
\label{pknu}
P_k(\nu)={q\over p}
\left[A_k(\nu)\!\!+\!\!\int_{\nu}^1\!\!\! dx {q+px\over x^p}
{[kp(1-x)]^{k-1}\over  k!}e^{-kp(1-x)}\right].
\end{equation}
The final distribution $P_k(\infty)$ can be written explicitly using
$\int_0^1 dx\, x^{l-1}(1-x)^{k-1}e^{\beta x} =B(l,k)\,
{}_1F_1(l,l+k,\beta)$, with $B(l,k)=\Gamma(l)\Gamma(k)/\Gamma(l+k)$ 
the Euler Beta function and ${}_1F_1(a,b,x)$ the confluent
hypergeometric function\cite{gr}.  Additionally, it is possible to
evaluate the large mass asymptotic behavior as the above integral is
dominated by a region of the order ${\cal O}(k^{-1})$ near the upper
limit.  Introducing the variable $y=xk$, keeping only the leading terms,
and performing the integration yields
\begin{equation}
\label{Pinfty}
P_k(\infty)\simeq {q^{1+p}\,\Gamma(q)\over \sqrt{2\pi p^2}}\,k^{-q-3/2}
e^{-k\Lambda(q)},
\end{equation}
with $\Lambda(q)$ as in Eq.~(\ref{qscl}). Regardless of $q$, the tail
of the distribution decays exponentially. However, in the limit $q\to
0$, a universal algebraic decay $P_k(\infty)\sim k^{-3/2}$ occurs in
the intermediate regime $k\ll q^{-2}$.  The average mass divergence
$q^{-1}$ agrees with Eq.~(\ref{pinf}).  Hence, the behavior is
fundamentally different than that found for mass independent rates as
the distribution of masses of passive clusters is characterized by a
finite scale.

The passive mass distribution (\ref{pknu}) can be re-cast into a
``scaling-like'' form. Using the same scaling variable $\xi=kq\nu$,
one finds
\begin{equation}
\label{Pkz}
P_k(t)\simeq P_k(\infty){\Gamma(q,\xi)\over \Gamma (q)}.
\end{equation}
This form involves the incomplete Gamma function, as was the case for
the function $G(\xi)$ in the case of constant kernel (see
Eq.~(\ref{psiphi})). Despite their very different nature, the mass
distributions of passive clusters in both constant and product kernel
models have a surprising similarity. Indeed, the former distribution
(\ref{pa}) can be written in the form (\ref{Pkz})  with $q$
replaced by $1+2/p$.

\section{Sum Kernel}

The aggregation process with a linear reaction rate, $K(i,j)=i+j$, has
been studied because of its mathematical simplicity as well as its
relevance to polymer formation\cite{ziff} and to shear-flow
coagulation \cite{drake}.  The sum kernel solution is described
concisely since it involves techniques similar to those used above.
The mass distributions evolve according to
\begin{eqnarray}
\label{mesum}
{d A_k\over dt}&=&{p\over 2}\,k\sum_{i+j=k}A_iA_j-A_k(kA+M),\nonumber\\
{d P_k\over dt}&=&{q\over 2}\,k\sum_{i+j=k}A_iA_j.
\end{eqnarray}
We introduce the modified densities ${\cal A}_k$ and the modified
time variable $\tau$ as follows
\begin{eqnarray}
\label{AA}
{\cal A}_k&=&A_k\,\exp\left[k\int_0^t dt'\,A(t')+\int_0^t dt'\,M(t')\right],
\nonumber\\
\tau&=&p\int_0^t dt_1\,\exp\left[-\int_0^{t_1} dt_2\,M(t_2)\right].
\end{eqnarray}
These transformations reduce Eq.~(\ref{mesum}) into
\begin{equation}
\label{aksum}
{d {\cal A}_k\over d\tau}={1\over 2}\,k\sum_{i+j=k} {\cal A}_i {\cal A}_j.
\end{equation}
The generating function ${\cal A}(z,\tau)=\sum_{k\geq 1} {\cal
  A}_k(\tau)\,e^{kz}$ evolves according to ${\partial {\cal A}\over
  \partial \tau} ={\cal A}\,{\partial {\cal A}\over \partial z}$, which
in turn can be re-written as ${\partial z\over \partial \tau}=-{\cal
  A}$.  Thus, for the monodisperse initial conditions, we get
\begin{equation}
\label{zAsum}
z=\ln {\cal A}-\tau{\cal A}.
\end{equation}

Again, the crucial part of the solution is the evaluation of the
number and mass densities $A(t)$ and $M(t)$.  These 
are related via the rate equation ${d\over dt}
A=-(1+q)AM$, as follows from Eqs.~(\ref{mesum}). Integrating over
time we obtain
\begin{equation}
\label{rmusum}
A^{1\over 1+q}=\exp\left[-\int_0^t  dt'\,M(t')\right].
\end{equation}
Inserting this equality into the definition of the generating function 
and using Eq.~(\ref{AA}), yields ${\cal A}=A^{q\over 1+q}$ for
$z=-\int_0^t dt'\,A(t')$.  Substituting this equality into
Eq.~(\ref{zAsum}) gives
\begin{equation}
\label{tAsum}
\ln A^{q\over 1+q}-\tau A^{q\over 1+q}=-\int_0^t dt'\,A(t').
\end{equation}
Since $\tau=p\int_0^t dt'\,A^{1\over 1+q}(t')$ can  be expressed
through $A(t)$, we conclude that Eq.~(\ref{tAsum}) is a closed equation
for $A(t)$.  It appears impossible, however, to find an explicit
solution for $A(t)$ or for $A(\tau)$.  Therefore, we proceed to
determine $\tau(A)$.  Differentiating Eq.~(\ref{tAsum}) with respect
to $A$ yields
\begin{equation}
\label{tauA}
{q\over 1+q}\left(A^{-1}-\tau A^{-{1\over 1+q}}\right)+
{q\over p}\,A^{q\over 1+q}{d\tau\over dA}=0.
\end{equation}
In deriving Eq.~(\ref{tauA}), we have applied the chain rule ${d\over
  dA}={d\tau\over dA}{dt\over d\tau}{d\over dt}$, and $\tau=p\int_0^t
dt'\,A^{1\over 1+q}(t')$ to evaluate the derivative of the right-hand
side of Eq.~(\ref{tAsum}).

We now multiply Eq.~(\ref{tauA}) by $A^{-{1\over 1+q}}$ and then
integrate to find an explicit expression for $\tau(A)$
\begin{equation}
\label{tA}
\tau=p\left[A^{-{q\over 1+q}}-A^{p\over 1+q}\right].
\end{equation}
One can verify that this expression is consistent with the expected
short time behaviors $A\cong 1-(1+q)t$ and $\tau\cong pt$ implied by
Eqs.~(\ref{mesum}) and (\ref{AA}).  In the long time limit, the above
expression yields the leading asymptotic behavior of the density and the
mass
\begin{equation}
\label{Amusum}
A(t)\simeq {q\over 1+q}\,t^{-1}, \qquad
M(t)\simeq {1\over 1+q}\,t^{-1}.
\end{equation}
Hence, these densities decay with the same universal exponent as in
the product kernel model. Again, the average active mass remains
finite $\langle k\rangle=M/A\to 1/q$ for $t\to \infty$. One slight
difference with the product kernel is that the average mass diverges in the
limit $q\to 0$.

To determine ${\cal A}_k$, we again employ the Lagrange inversion
formula. {}From the exponentiated form of Eq.~(\ref{zAsum}), ${\cal
  A}e^{-\tau{\cal A}}=e^z$, we immediately find ${\cal
  A}_k=(k\tau)^{k-1}/k!$, or \hbox{$A_k(t)=(k\tau)^{k-1}\exp\{-\int_0^t
  dt'[kA(t')+M(t')]\}/k!$}.  Using Eqs.~(\ref{tAsum}) and (\ref{tA}),
the mass distribution of active clusters is obtained in terms of the
variable $\nu=A^{1\over 1+q}$
\begin{equation}
\label{Aktnew}
A_k(\nu)=\nu^{1+q}\,{[kp(1-\nu)]^{k-1}\over k!}\,e^{-kp(1-\nu)}.
\end{equation}
Although the variable $\nu$ is quite different, the similarity with
the corresponding product kernel expression (\ref{Akmu}) is striking.
The monomer density in this case is $A_1=A\exp[-p(1-A^{1/(1+q)})]$.
Therefore, it follows the same $t^{-1}$ asymptotic decay as in the
product kernel case.

The appropriate large mass and large time limit is again taken in such a
way, that the scaling variable $\xi=kq\nu$ is kept finite. The
corresponding characteristic size grows according to $k^*\sim
t^{1/(1+q)}$. The mass distribution of active clusters can again be
re-casted into the following scaling-like form
\begin{equation}
\label{qsclsum}
A_k(t)\sim e^{-\Lambda(q) k}\,k^{-3/2}\,t^{-1}\,F(\xi),
\end{equation}
with $\Lambda(q)=-q-\ln(1-q)$, and with the scaling function
$F(\xi)=e^{-\xi}$. The numerous similarities between corresponding
expressions in the product and sum kernel solutions reflect the
similar underlying mathematical structure of both models. For example,
the distributions generally decay exponentially. Furthermore, in the
limit $q\to 0$, there is an intermediate regime $k\ll q^{-2}$, where
an algebraic decay $A_k(t)\sim k^{-3/2}$ occurs.

We analyzed the mass distribution of passive clusters along the same
lines as for the product kernel model. For example, the final distribution 
is given by 
\begin{equation}
\label{pfinalsum}
P_k(\infty)={q\over p}\int_0^1\!\!\! dx x^q[(kq\!+\!1)\!+\!kpx]
{[kp(1\!-\!x)]^{k-1}\over k!}e^{-kp(1-x)},
\end{equation}
which may be rewritten explicitly through the beta function and the
confluent hypergeometric function.  Interestingly, the leading large
mass behavior of the final distribution is {\it identical} to the one
obtained for the product kernel case (\ref{Pinfty}). Additionally, the
time dependent behavior can be rewritten using the form (\ref{Pkz})
with the argument of the Gamma function now being $1+q$ (instead of
$q$).

\section{Discussion}

We have analyzed a class of stochastic aggregation processes.  We have
solved for the three classical reaction kernels: $1, i+j$, and $ij$.
Our methods are also applicable to the linear combination
$K(i,j)=a+b(i+j)+cij$, with $a,b$, and $c$ non-negative
constants\cite{spouge,treat}, and to the reaction kernel
$K(i,j)=2-r^{i}-r^{j}$, where $0<r<1$ \cite{leyvraz}.

We have observed that stochastic aggregation exhibits a variety of
universal features.  For instance, the final number density of passive
clusters is identical in all three models.  Also, the number density
of active clusters decays as $t^{-1}$. The former feature has been
explained by a simple argument; remarkably, it remains valid even
beyond the scope of the rate equations.  It would be interesting to
understand the latter feature within a coherent model-independent
framework. For the constant and sum kernels, we have seen that both
the number density, the monomer density, and the mass density decay  
according to $t^{-1}$. Furthermore, the results suggest that the
following scaling-like form
\begin{equation}
\label{hypothesis}
A_k(t)\simeq \rho_k\,t^{-1}\,F(\xi),
\end{equation}
applies when the kernel grows indefinitely with the mass. For the
solvable sum and product kernels, $\rho_k$ decays exponentially in the
large mass limit.  It will be interesting to study whether this
exponential behavior applies for more general kernels, e.g., for
homogeneous kernels, $K(ai,aj)=a^\lambda K(i,j)$, with a positive
homogeneity index $\lambda>0$.

The simplest constant kernel model is special, and it exhibits the
most interesting behavior.  We have shown that in this case, the final
mass distribution of passive clusters is algebraic. The decay exponent
is governed by the  probability $p$, and the entire range
of decay exponents, consistent with mass conservation is possible. The
time dependent behavior follows ordinary scaling, and the
corresponding mass scale grows algebraically with a nonuniversal
($p$-dependent) exponent.

Very similar behaviors were found for the product and sum kernels. In
general, the mass distributions decay exponentially, and the average
mass approaches a finite value. In the limit $q\to 0$, corresponding
to the original aggregation process, an algebraic mass decay occurs in
the intermediate mass range $k\ll q^{-2}$. The time dependent behavior
does not follow ordinary scaling, as the distributions are dominated
by time independent exponential factors. Nevertheless, a corrective
factor which is governed by an algebraically growing scale was found
to be relevant. Remarkably, although the models exhibit different
temporal behaviors, they possess the same large mass behavior of the
final  mass distribution.

The above stochastic aggregation provides an interpolation between
aggregation ($p=1$) and annihilation ($p=0$) processes. Therefore, it
may be used as a tool for studying either problems. For example, the
Smoluchowski aggregation equations may be applied to low dimensional
processes using effective reaction rates. Hence, the above solution of
the master equations may be relevant to reaction-diffusion problems.

\medskip\noindent
This research was supported by the DOE (W-7405-ENG-36), NSF
(DMR9632059), and ARO (DAAH04-96-1-0114).

\end{multicols}
\end{document}